\documentclass{revtex4}
\usepackage{amsmath,amsthm}
\usepackage{rotating}
\usepackage{blkarray}
\usepackage{booktabs}
\usepackage{mathtools}
\usepackage{multirow}
\usepackage{adjustbox}

\usepackage{graphicx}

\usepackage{amsfonts}
\newtheorem{theorem}{Theorem}[section]
\newtheorem{lemma}[theorem]{Lemma}

\theoremstyle{remark}
\newtheorem{remark}[theorem]{Remark}

\theoremstyle{definition}

\theoremstyle{example}
\newtheorem{example}[theorem]{Example}

\theoremstyle{notation}

\newcommand{\bra}[1]{\langle#1|}
\newcommand{\ket}[1]{|#1\rangle}

\begin{document}

\title{Fast Fourier transforms and fast Wigner and Weyl functions in large quantum systems}            
\author{C. Lei and A. Vourdas}
\affiliation{Department of Computer Science,\\
University of Bradford, \\
Bradford BD7 1DP, United Kingdom\\c.lei1@bradford.ac.uk\\a.vourdas@bradford.ac.uk}

\begin{abstract}
Two methods for fast Fourier transforms are used in a quantum context.
The first method is for systems with dimension of the Hilbert space $D=d^n$ with $d$ an odd integer, and is inspired by the Cooley-Tukey formalism.
The `large Fourier transform' is expressed as a sequence of $n$ `small Fourier transforms' (together with some other transforms) in quantum systems with $d$-dimensional Hilbert space.
Limitations of the method are discussed.
In some special cases, the $n$ Fourier transforms can be performed in parallel.
The second method is for systems with dimension of the Hilbert space $D=d_0...d_{n-1}$ with $d_0,...,d_{n-1}$ odd integers coprime to each other.
It is inspired by the Good formalism, which in turn is based on the Chinese reminder theorem.
In this case also the `large Fourier transform' is expressed as a sequence of $n$ `small Fourier transforms' (that involve some constants related to the number theory that describes the formalism).
The `small Fourier transforms' 
can be performed  in a classical computer or in a quantum computer (in which case we have the additional well known advantages of quantum Fourier transform circuits).
In the case that the small Fourier transforms are performed with a classical computer, complexity arguments for both methods show the reduction in computational time from ${\cal O}(D^2)$ to ${\cal O}(D\log D)$.
The second method is also used for the fast calculation of Wigner and Weyl functions, in quantum systems with large finite dimension of the Hilbert space.
\end{abstract}
\maketitle

\section{Introduction}
The fast implementation of large Fourier transforms is very important for many technological applications.
Roughly speaking in this paper we express the Fourier transform in a Hilbert space of large dimension, as a combination of many Fourier transforms in Hilbert spaces of small dimension.
This is a fast Fourier transform, because performing many `small' Fourier transforms instead of one  `large' Fourier transform, is computationally beneficial.
The `small' Fourier transforms can be performed  in a classical computer or as quantum Fourier transforms in a quantum computer.
In the latter case, we will also have
an additional well known reduction of the computational time by quantum Fourier transform circuits (e.g., \cite{NC,PF}).
Our methodology (and the associated reduction of computational time) is applicable to the calculation of other quantities also, like the Wigner and Weyl functions.

Two important approaches are the Cooley-Tukey formalism \cite{B0,B00}, and the Good formalism\cite{G1,G2,G3}  which is based on the Chinese remainder theorem.
There are also many variations of these schemes (reviewed in \cite{B1,B2,B3}).
In this paper we study the implementation of fast Fourier transforms in quantum systems with large dimension of the Hilbert space.
We also study the fast calculation of the Wigner and Weyl functions.
This is an important application of the physics of quantum systems with finite -dimensional Hilbert space(e.g. \cite{V1}).

 We consider a finite quantum system $\Sigma (D)$ with variables in ${\mathbb Z}(D)$ (the ring of integers modulo $d$) where $D$ is an odd integer.
 This system is described by the $D$-dimensional Hilbert space $H(D)$.
 There are well known technical differences between quantum systems with odd dimension $D$ and even dimension $D$ (e.g., \cite{EV0,EV1,EV2}).
In this paper we consider systems with odd dimension $D$.
We discuss the fast implementation of the Fourier transform $F$ in $\Sigma (D)$, using two methods described briefly below.

\subsection{First method for the case $D=d^n$ with $d$ an odd integer}
The fast implementation of the Fourier transform $F$ in $\Sigma (D)$, is using a sequence of $n$ Fourier transforms (together with some other transforms) in 
a multipartite system $\Sigma_n(d)$ comprised of $n$ components each of which is described with variables in ${\mathbb Z}(d)$.
Positions and momenta in $\Sigma_n(d)$ take values in $[{\mathbb Z}(d)]^n={\mathbb Z}(d)\times ...{\mathbb Z}(d)$ and the corresponding  Hilbert space is ${\mathfrak H}_A=H(d)\otimes ...\otimes H(d)$.
The Hilbert spaces $H(D)$ and ${\mathfrak H}_A$ are isomorphic (they have the same dimension), and in this sense $\Sigma (D)$ and $\Sigma_n(d)$ are two different descriptions of the same system.
However, Fourier transforms and other phase space methods are different in these two cases\cite{LV}.

Mathematically, this approach  is inspired by the 
Cooley-Tukey formalism \cite{B0} for fast Fourier transforms (see also \cite{B1,B2,B3}), and is used here in a quantum context.
But we note that the most popular Cooley-Tukey algorithm is for $D=2^n$, whilst in our approach $D$ is a power of an odd number. 

\begin{figure}
\centering
\includegraphics[scale =0.4]{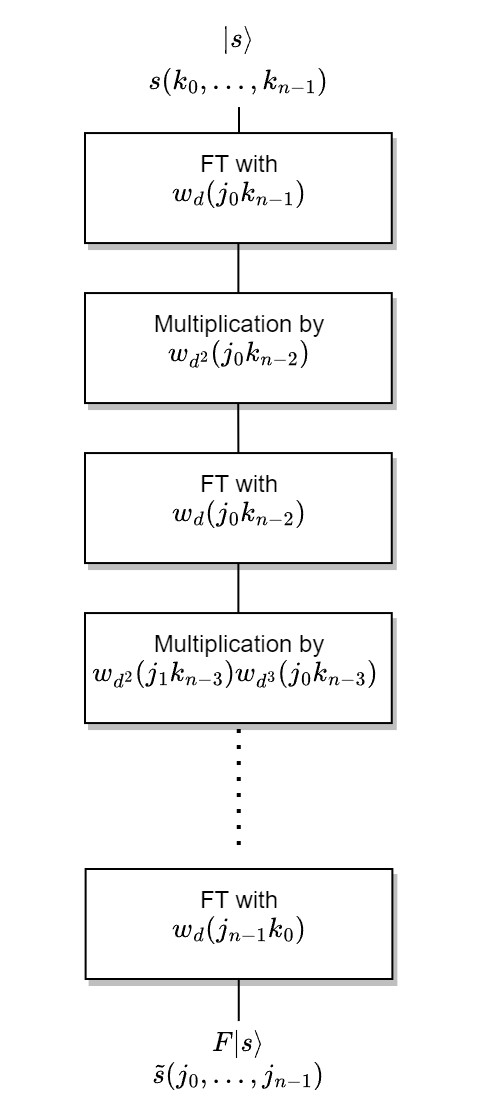}
 \caption{Circuit for the fast calculation of the Fourier transform $F$, using Eqs(\ref{20})-(\ref{30}). Here $D=d^n$.}

\label{Fig1}
\end{figure}

A quantum circuit for the implementation of this fast Fourier transform is given in Fig.\ref{Fig1}.
In some special cases, the various operations  can be performed in parallel (parallel computing).

We discuss the complexity of this method and show that the computational time is reduced from ${\cal O}(D^2)$ to ${\cal O}(D\log D)$.
We also present numerical work that supports this.

A limitation of the method is the fact that the ring ${\mathbb Z}(D)$ (with $D=d^n$) is not isomorphic to the ring $[{\mathbb Z}(d)]^n$.
Although there is a bijective map between them, sum and products do not correspond to sums and products (section 2.A).
The implications of this are discussed in section 4.C.
For example, this method cannot be used in Eqs(\ref{WW}) below, for the fast calculation of the Wigner and Weyl functions.

\subsection{Second method for the case $D=d_0...d_{n-1}$ with $d_0,...,d_{n-1}$ odd integers coprime to each other}
 The fast implementation of the Fourier transform F, is using a
multipartite system $\Sigma(d_0,...,d_{n-1})$ comprised of $n$ components, which are described with variables in ${\mathbb Z}(d_0),...,{\mathbb Z}(d_{n-1})$.
Positions and momenta in $\Sigma(d_0,...,d_{n-1})$ take values in ${\mathbb Z}(d_0)\times...\times {\mathbb Z}(d_{n-1})$ and the corresponding  Hilbert space is ${\mathfrak H}_B=H(d_0)\otimes ...\otimes H(d_{n-1})$.
The Hilbert spaces $H(D)$ and ${\mathfrak H}_B$ are isomorphic (they have the same dimension), and in this sense $\Sigma (D)$ and  $\Sigma(d_0,...,d_{n-1})$  are two different descriptions of the same system.

Mathematically, this approach  is inspired by the 
Good formalism\cite{G1,G2,G3} for fast Fourier transforms (see also \cite{B1,B2,B3}), which in turn is based on the Chinese remainder theorem, and is used here in a quantum context.
A quantum circuit for the implementation of this fast Fourier transform is given in Fig.\ref{Fig5}.

The complexity of the method is discussed, and it is shown that the computational time is reduced from ${\cal O}(D^2)$ to ${\cal O}(D\log D)$.
This is supported with numerical work.

A strength of the method is the fact that the ring ${\mathbb Z}(D)$ is isomorphic to the ring ${\mathbb Z}(d_0)\times...\times {\mathbb Z}(d_{n-1})$ (section 2.B).
Because of this the method is used for the fact calculation of Wigner and Weyl functions in section 6.

\subsection{Contents }
The work is complementary to the work on quantum Fourier transforms.
It reduces a large Fourier transform to many small Fourier transforms, and this reduces the computational time.
The small Fourier transforms can be preformed with a classical computer, or (if available) with a quantum computer so that we have the additional (and well known) advantages of
quantum Fourier transforms\cite{NC,PF}.

In section 2 we discuss the number theory related to the two methods.
 In section 3 we consider a quantum system $\Sigma(D)$ with variables in ${\mathbb Z}(D)$ where $D$ is an odd integer, described by the $D$-dimensional Hilbert space $H(D)$.
In section 4 we present the first method for the case where $D=d^n$.
In section 5 we present the second method for the case where $D=d_0...d_{n-1}$ with $d_0,...,d_{n-1}$ odd integers coprime to each other. 
In section 6 we use the second method for the fast calculation of the Wigner and Weyl functions.
We conclude in section 7 with a discussion of our results.

\section{Number theory for the two fast Fourier transforms}
\subsection{A bijective map between the non-isomorphic rings $[{\mathbb Z}(d)]^n$ and ${\mathbb Z}(D)$ when $D=d^n$}
${\mathbb Z}(D)$ is the ring of integers modulo $D$, where $D$ is an odd integer. 
We take $D=d^n$ (where $d$ is an odd integer) and 
consider a bijective map between $[{\mathbb Z}(d)]^n={\mathbb Z}(d)\times...\times{\mathbb Z}(d)$ and ${\mathbb Z}(D)$. We use upper case letters for elements in ${\mathbb Z}(D)$,
and lower case letters for elements in ${\mathbb Z}(d)$.
We also take $j_r\in {\mathbb Z}(d)$ and $J\in {\mathbb Z}(D)$ in the `periods' 
\begin{eqnarray}\label{47}
\left [-\frac{d-1}{2}, \frac{d-1}{2}\right];\;\;\;
\left [-\frac{D-1}{2}, \frac{D-1}{2}\right],
\end{eqnarray}
correspondingly.

We introduce the following bijective map between the sets $[{\mathbb Z}(d)]^n$ and ${\mathbb Z}(D)$
\begin{eqnarray}\label{16}
(j_0,...,j_{n-1})\;\leftrightarrow\;J=j_0+j_1d+...+j_{n-1}d^{n-1}.
\end{eqnarray}
Given $J$, we can find the $j_0,...,j_{n-1}$ as the remainders in the following sequence of divisions:
\begin{itemize}
\item
We divide $J$ by $d$ and we get $j_1+j_2d+...j_{n-1}d^{n-2}$ and remainder $j_0$.
\item
We divide $j_1+j_2d+...j_{n-1}d^{n-2}$ by $d$ and we get $j_2+j_3d+...j_{n-1}d^{n-3}$ and remainder $j_1$.
\item
e.t.c.
\end{itemize}
We note that 
the $[{\mathbb Z}(d)]^n$ as a ring (with addition and multiplication componentwise), is not isomorphic to the ring ${\mathbb Z}(D)$ because addition and multiplication is different\cite{LV}. Indeed
\begin{eqnarray}
(j_0,...,j_{n-1})+(k_0,...,k_{n-1})=(j_0+k_0,...,j_{n-1}+k_{n-1})
\end{eqnarray}
does not correspond to $J+K$.
The sum in ${\mathbb Z}(D)$ has  the `carry' rule and the $r$-component might be $j_r+k_r+1$ rather than $j_r+k_r$ .
In contrast, there is no `carry' rule in $[{\mathbb Z}(d)]^n$.
Also the multiplication in  ${\mathbb Z}(D)$ 
\begin{eqnarray}\label{P9}
JK=j_0k_0+d(j_1k_0+k_1j_0)+...+d^{n-1}(j_0k_{n-1}+...+j_{n-1}k_0),
\end{eqnarray}
does not correspond to the componentwise multiplication in $[{\mathbb Z}(d)]^n$
\begin{eqnarray}
(j_0,...,j_{n-1})\cdot (k_0,...,k_{n-1})=(j_0k_0,...,j_{n-1}k_{n-1}).
\end{eqnarray}
Due to the non-isomorphism of the rings ${\mathbb Z}(D)$ and $[{\mathbb Z}(d)]^n$, there is a limitation (see subsection 4.C) of the corresponding fast Fourier transform method in section 4.

We use the notation
\begin{eqnarray}
\omega_r(s)=\exp \left (i\frac{2\pi s}{r}\right).
\end{eqnarray}
For later use, we use Eq(\ref{P9}) and we get
\begin{eqnarray}\label{56}
\omega_D(JK)=\omega_{d^n}(j_0k_0)\omega_{d^{n-1}}(j_1k_0+k_1j_0)...\omega_d(j_0k_{n-1}+...+j_{n-1}k_0).
\end{eqnarray}

\begin{example}
We consider the bijective map between the sets $[{\mathbb Z}(3)]^2$ and ${\mathbb Z}(9)$:
\begin{eqnarray}
(j_0,j_1)\;\leftrightarrow\;J=j_0+3j_1;\;\;\;j_\nu=-1,0,1;\;\;\;J=-4,...,4.
\end{eqnarray}
Then $(1,1)$ corresponds to $4\in {\mathbb Z}(9)$.
Addition in $[{\mathbb Z}(3)]^2$ gives $(1,1)+(1,1)=(-1,-1)$ which corresponds to $-4\in {\mathbb Z}(9)$.
The corresponding addition in ${\mathbb Z}(9)$ gives $4+4=-1$.
\end{example}

\subsection{The isomorphic rings ${\mathbb Z}(d_0)\times ...\times {\mathbb Z}(d_{n-1})$ and ${\mathbb Z}(D)$ when $D=d_0...d_{n-1}$ and the $d_0,...,d_{n-1}$ are coprime}

A different method for Fast Fourier transforms is the Good method \cite{G1,G2,G3} which is based on the Chinese remainder theorem.
In a quantum context it has been used in \cite{V3,V1}.

If $d_0,...,d_{n-1}$ are coprime, then the ring ${\mathbb Z}(d_0)\times ...\times {\mathbb Z}(d_{n-1})$ is isomorphic to ${\mathbb Z}(D)$ where $ D=d_0\times...\times d_{n-1}$.
We first define the integers
\begin{eqnarray}\label{7}
{\mathfrak a}_\nu=\frac{D}{d_\nu};\;\;\;{\mathfrak a}_\nu {\mathfrak b}_\nu=1 ({\rm mod}\; d_\nu)
\end{eqnarray}
${\mathfrak b}_\nu$ is the inverse of ${\mathfrak a}_\nu$ within ${\mathbb Z}(d_\nu)$, and it exists because the ${\mathfrak a}_\nu, d_\nu$ are coprime.
We also define the ${\mathfrak c}_\nu={\mathfrak a}_\nu {\mathfrak b}_\nu$ as an element of ${\mathbb Z}(D)$, which is an integer multiple of $d_\nu$ plus one ($Nd_\nu+1$).
\begin{lemma}
\begin{eqnarray}\label{10}
{\mathfrak a}_\nu {\mathfrak a}_\mu={\mathfrak a}_\nu^2\delta_{\mu \nu}({\rm mod}\;D);\;\;\;{\mathfrak c}_\nu {\mathfrak c}_\mu={\mathfrak c}_\nu\delta_{\mu \nu}({\rm mod}\;D);\;\;\;
{\mathfrak a}_\nu {\mathfrak c}_\mu={\mathfrak a}_\nu\delta_{\mu \nu}({\rm mod}\;D).
\end{eqnarray}
\end{lemma}

\begin{proof}
In the first relation, for $\nu\ne \mu$ we get a multiple of $D$, which is $0({\rm mod}\; D$).

In the second relation, we get
\begin{eqnarray}\label{10}
{\mathfrak c}_\nu {\mathfrak c}_\mu&=&{\mathfrak a}_\nu {\mathfrak b}_\nu {\mathfrak a}_\mu {\mathfrak b}_\mu=({\mathfrak a}_\nu {\mathfrak b}_\nu)^2 \delta_{\nu\mu}=
{\mathfrak c}_\nu ^2\delta_{\nu \mu}={\mathfrak c}_\nu (N d_\nu +1)\delta_{\nu\mu}={\mathfrak c}_\nu \delta_{\nu\mu}+N{\mathfrak b}_\nu ({\mathfrak a}_\nu d_\nu)
\nonumber\\&=&{\mathfrak c}_\nu \delta_{\nu\mu}+N{\mathfrak b}_\nu D={\mathfrak c}_\nu \delta_{\nu\mu}\;({\rm mod}\;D).
\end{eqnarray}

In the third relation, we get
\begin{eqnarray}\label{10}
{\mathfrak a}_\nu {\mathfrak c}_\mu&=&{\mathfrak a}_\nu  {\mathfrak a}_\mu {\mathfrak b}_\mu={\mathfrak a}_\nu^2 {\mathfrak b}_\nu \delta_{\nu\mu}=
{\mathfrak a}_\nu {\mathfrak c}_\nu \delta_{\nu \mu}={\mathfrak a}_\nu (N d_\nu +1)\delta_{\nu\mu}={\mathfrak a}_\nu \delta_{\nu\mu}+ND
\nonumber\\&=&{\mathfrak c}_\nu \delta_{\nu\mu}\;({\rm mod}\;D).
\end{eqnarray}

\end{proof}

We define a bijective map between  ${\mathbb Z}(d_1)\times ...\times {\mathbb Z}(d_n)$ and ${\mathbb Z}(D)$ as follows:
\begin{eqnarray}\label{9}
(j_0,...,j_{n-1})\;\leftrightarrow\;J;\;\;\;j_\nu=J{(\rm mod}\;d_\nu)\in {\mathbb Z}(d_\nu);\;\;\;J=\sum j_\nu {\mathfrak c}_\nu\in {\mathbb Z}(D).
\end{eqnarray}
The Chinese remainder theorem ensures that this map is bijective.
Using Eq.(\ref{10}), we prove  that
\begin{eqnarray}\label{170}
&&(j_0+j_0' ,...,j_{n-1}+j_{n-1}')\;\leftrightarrow\;J+J';\nonumber\\
&&(j_0j_0',...,j_{n-1}j_{n-1}')\;\leftrightarrow\;JJ'.
\end{eqnarray}
and therefore the ring ${\mathbb Z}(d_0)\times ...\times {\mathbb Z}(d_{n-1})$ is isomorphic to ${\mathbb Z}(D)$.

We also define a different bijective map 
\begin{eqnarray}\label{90}
(\widehat  j_0,...,\widehat j_{n-1})\;\leftrightarrow\;J;\;\;\;\widehat j_\nu=J{\mathfrak b}_\nu {(\rm mod}\;d_\nu)\in {\mathbb Z}(d_\nu);\;\;\;J=\sum\widehat j_\nu {\mathfrak a}_\nu\in {\mathbb Z}(D).
\end{eqnarray}
From Eqs(\ref{9}),(\ref{90}) we find the relationship between $j_\nu$ and $\widehat j_\nu$:
\begin{eqnarray}\label{QQ}
\widehat j_\nu=j_\nu {\mathfrak b}_\nu {(\rm mod}\;d_\nu);\;\;\;j_\nu=\widehat j_\nu {\mathfrak a}_\nu {(\rm mod}\;d_\nu).
\end{eqnarray}
Using  Eqs(\ref{10}),(\ref{9}),(\ref{90}) we prove that
\begin{eqnarray}\label{GG}
JK=\widehat j_0k_0{\mathfrak a}_0+...+\widehat j_{n-1} k_{n-1}{\mathfrak a}_{n-1}.
\end{eqnarray}
It then follows the important relation:
\begin{eqnarray}\label{13A}
\omega_D(JK)&=&\omega_{d_0}(\widehat j_0k_0)...\omega_{d_{n-1}}(\widehat j_{n-1}k_{n-1})\nonumber\\
&=&\omega_{d_0}(j_0{\mathfrak b}_0k_0)...\omega_{d_{n-1}}(j_{n-1}{\mathfrak b}_{n-1}k_{n-1})
\end{eqnarray}

\begin{example}
Let $d_0=3$ and $d_1=5$. Then $D=15$ and
\begin{eqnarray}
&&{\mathfrak a}_0=5;\;\;\;{\mathfrak b}_0=2;\;\;\;{\mathfrak c}_0=10\nonumber\\
&&{\mathfrak a}_1=3;\;\;\;{\mathfrak b}_1=2;\;\;\;{\mathfrak c}_1=6.
\end{eqnarray}
Then 
\begin{eqnarray}
J=10j_0+6j_1=5\widehat j_0+3\widehat j_1
\end{eqnarray}
As an example, we take $J=11$ and we find the corresponding $(j_0,j_1)=(2,1)$ and $(\widehat j_0,\widehat j_1)=(4,2)$.
We confirm Eq(\ref{QQ}):
\begin{eqnarray}
&&j_0 {\mathfrak b}_0 =2\times 2=4=\widehat j_0;\;\;\;j_1 {\mathfrak b}_1 =1\times 2=\widehat j_1\nonumber\\
&&\widehat j_0 {\mathfrak a}_0=4\times 5=2 {(\rm mod}\;3)=j_0;\;\;\;\widehat j_1 {\mathfrak a}_1=2\times 3=1 {(\rm mod}\;5)=j_1.
\end{eqnarray}

\end{example}

\section{A quantum system $\Sigma(D)$ with variables in ${\mathbb Z}(D)$ }

We  consider a quantum system $\Sigma(D)$ with variables in  the ring ${\mathbb Z}(D)$, where $D$ is an odd integer. 
$H(D)$ is the $D$-dimensional Hilbert space describing this system.

Let $|X;J\rangle$ where $J\in {\mathbb Z}(D)$ be an orthonormal basis in $H(D)$.
The $X$ in the notation is not a variable, it simply indicates `position states'.
The finite Fourier transform $F$ is given by\cite{T}
\begin{eqnarray}\label{FF}
&&F=\frac{1}{\sqrt{D}}\sum _{J,K}\omega_D(JK) \ket{X;J}\bra{X;K};\;\;\;A,J,K\in{\mathbb Z}(D)\nonumber\\
&&F^4={\bf 1};\;\;\;FF^{\dagger}={\bf 1}.
\end{eqnarray}

We act with $F^\dagger$ on position states and get the dual  basis
\begin{eqnarray}
&&\ket{P;J}=F^\dagger \ket{X;J}=\frac{1}{\sqrt{D}}\sum _K\omega_D(-JK)\ket{X;K}.
\end{eqnarray}
The $P$ in the notation is not a variable, it simply indicates `momentum states'.
A state $\ket{s}$ in $H(D)$ can be written as
\begin{eqnarray}\label{TR}
&&\ket{s}=\sum s(J)\ket{X;J}=\sum {\widetilde s}(J)\ket{P;J}\nonumber\\
&&{\widetilde s}(J)=\frac{1}{\sqrt{D}}\sum _K\omega_D(JK)s(K)
\end{eqnarray}
Below we study the fast implementation of this Fourier transform.

\section{The case $D=d^n$}

\subsection{A multipartite system $\Sigma_n(d)$ with variables in $[{\mathbb Z}(d)]^n$ }
We consider a multipartite system $\Sigma_n(d)$ comprised of $n$ components each of which is described with variables in ${\mathbb Z}(d)$.
Positions and momenta take values in $[{\mathbb Z}(d)]^n$. 
This system is described with the $d^n$-dimensional Hilbert space ${\mathfrak H}_A=H(d)\otimes ...\otimes H(d)$.
We consider the basis
\begin{eqnarray}
\ket{X;j_0,...,j_{n-1}}=\ket{X;j_0}\otimes...\otimes\ket{X;j_{n-1}};\;\;\;j_r\in{\mathbb Z}(d).
\end{eqnarray}
An arbitrary state is written as
\begin{eqnarray}\label{5}
\ket{s}=\sum s(j_0,...,j_{n-1})\ket{X;j_0,...,j_{n-1}};\;\;\;\sum |s(j_0,...,j_{n-1})|^2=1.
\end{eqnarray}

Fourier transforms are defined as:
\begin{eqnarray}
&&{\mathfrak F}_A={\cal F}\otimes ...\otimes {\cal F};\;\;\;{\cal F}=\frac{1}{\sqrt{d}}\sum _{j,k}\omega_d(jk) \ket{X;j}\bra{X;k}\nonumber\\
&&{\mathfrak F}_A^4={\bf 1};\;\;\;{\mathfrak F}_A{\mathfrak F}_A^{\dagger}={\bf 1};\;\;\;j,k\in{\mathbb Z}(d).
\end{eqnarray}

We assume that $D=d^n$ and compare and contrast the systems $\Sigma_n(d)$ and $\Sigma(D)$.
Then ${\mathfrak H}_A$ is isomorphic to $H(D)$ (because they both have the same dimension), and therefore $\Sigma (D)$ and $\Sigma_n(d)$ are two different descriptions of the same system.
However as discussed in ref\cite{LV}, Fourier transforms and phase space methods (displacement operators, Wigner and Weyl functions, etc) are different in  $\Sigma(D)$ and $\Sigma_n(d)$ 
($F$ is different from ${\mathfrak F}_A$).
This is because in these techniques we use addition and multiplication and as we explained above, the rings $[{\mathbb Z}(d)]^n$ and ${\mathbb Z}(D)$ are not isomorphic to each other.
Furthermore (proposition 4.4 in ref\cite{LV}), depending on the $d,n$, the Fourier transforms in $\Sigma_n(d)$ and $\Sigma(D)$ are unitarily inequivalent or unitarily equivalent .

Below we explain how the equivalent of the Fourier transform $F$ in $\Sigma(D)$, is a sequence of transformations in $\Sigma_n(d)$ that involve Fourier transforms in the various components together with some other transformations.
The latter is a fast Fourier transform in a quantum context.

\subsection{Fast Fourier transform $F$ in $\Sigma(D)$ as a sequence of transformations in $\Sigma_n(d)$ with $D=d^n$}\label{sec10}

We use the following  dual notation for functions and states in $\Sigma(D)$, based on the bijective map in Eq.(\ref{16}):
\begin{eqnarray}\label{234}
s(K)=s( k_0,...,k_{n-1}).
\end{eqnarray}

The matrix elements of the Fourier transform $F$ in $\Sigma(D)$ (Eq.(\ref{FF})) as:
\begin{eqnarray}\label{99A}
 &&F(j_0,...,j_{n-1}|k_0,...,k_{n-1})=\bra{j_0,...,j_{n-1}}F\ket{k_0,...,k_{n-1}}\nonumber\\
  &&=\frac{1}{\sqrt{d^n}}\omega _{d^n}[j_0k_0+d(j_1k_0+k_1j_0)+...+d^{n-1}(j_0k_{n-1}+...+j_{n-1}k_0)]\nonumber\\
  &&={\cal A}(k_{n-1}){\cal A}(k_{n-2}){\cal A}(k_{n-3})... {\cal A}(k_0)
 \end{eqnarray}
 where
 \begin{eqnarray}\label{25}
 &&{\cal A}(k_{n-1})=\frac{1}{\sqrt{d}}\omega_d(j_0k_{n-1})\nonumber\\
 &&{\cal A}(k_{n-2})=\frac{1}{\sqrt{d}}\omega_{d}(j_1k_{n-2}) \omega_{d^2}(j_0k_{n-2})\nonumber\\ 
 &&{\cal A}(k_{n-3})=\frac{1}{\sqrt{d}}\omega_{d}(j_2k_{n-3})\omega_{d^2}(j_1k_{n-3}) \omega_{d^3}(j_0k_{n-3})\nonumber\\ 
 &&......\nonumber\\
 &&{\cal A}(k_0)=\frac{1}{\sqrt{d}}\omega_{d}(j_{n-1}k_{0})\omega_{d^2}(j_{n-2}k_0).... \omega_{d^n}(j_0k_0)
 \end{eqnarray}
 Using this we implement the Fourier transform in Eq.(\ref{TR}), as a sequence of transforms in the system $\Sigma_n(d)$.
 It involves the following steps (shown also in the quantum circuit in Fig.\ref{Fig1}):
  \begin{itemize}
 \item
 A Fourier transform of $s(K)=s(k_0,...,k_{n-1})$ with $\omega_d(j_0k_{n-1})$ that involves summation over $k_{n-1}$:
 \begin{eqnarray}\label{20}
 s_1(j_0|k_0,..,k_{n-2})=\frac{1}{\sqrt{d}}\sum _{k_{n-1}}\omega_d(j_0k_{n-1})s(k_0,...,k_{n-1})
   \end{eqnarray}

 \item
 We first multiply $s_1(j_0|k_0,..,k_{n-2})$  by $\omega_{d^2}(j_0k_{n-2})$ (this is the analogue of `twiddle factors'\cite{GS} in the present context).
  Then we perform a Fourier transform of $\omega_{d^2}(j_0k_{n-2})s_1(j_0|k_0,..,k_{n-2})$ with $\omega_{d}(j_1k_{n-2})$,  that involves summation over $k_{n-2}$:
 \begin{eqnarray}\label{20A}
 s_2(j_0,j_1|k_0,..,k_{n-3})&=&\frac{1}{\sqrt{d}}\sum _{k_{n-2}}\omega_{d}(j_1k_{n-2}) [\omega_{d^2}(j_0k_{n-2})s_1(j_0|k_0,..,k_{n-2})]
  \end{eqnarray}

 \item
 We first multiply $s_2(j_0,j_1|k_0,..,k_{n-2})$ by $\omega_{d^2}(j_1k_{n-3})\omega_{d^3}(j_0k_{n-3})$.
  Then we perform a Fourier transform of $\omega_{d^2}(j_1k_{n-3})\omega_{d^3}(j_0k_{n-3})s_2(j_0,j_1|k_0,..,k_{n-2})$ with $\omega_{d}(j_2k_{n-3})$, that involves summation over $k_{n-3}$:
 \begin{eqnarray}
 s_3(j_0,j_1,j_2|k_0,..,k_{n-4})&=&\frac{1}{\sqrt{d}}\sum _{k_{n-3}}\omega_{d}(j_2k_{n-3}) [\omega_{d^2}(j_1k_{n-3})\omega_{d^3}(j_0k_{n-3})s_2(j_0,j_1|k_0,..,k_{n-2})]
  \end{eqnarray}

   \item
  We continue in this way and the $n$-step is  a Fourier transform with $\omega_{d}(j_{n-1}k_{0})$ that involves summation over $k_{0}$:
 \begin{eqnarray}\label{30}
{\widetilde s}(J)= {\widetilde s}(j_0,...,j_{n-1})&=&\frac{1}{\sqrt{d}}\sum _{k_0}\omega_{d}(j_{n-1}k_{0})[\omega_{d^2}(j_{n-2}k_0).... \omega_{d^n}(j_0k_0)s_{n-1}(j_0,...,j_{n-2}|k_0)]
  \end{eqnarray}
 
 \end{itemize}
 We note that:
 \begin{itemize}
 \item
 \begin{eqnarray}
 \sum |s(k_0,...,k_{n-1})|^2=\sum |s_1(j_0|k_0,..,k_{n-2})|^2=...=\sum|{\widetilde s}(j_0,...,j_{n-1})|^2=1.
 \end{eqnarray} 
 \item
  Starting from $s_r(j_0,...,j_{r-1}|k_0,..,k_{n-r-1})$ with a series of inverse Fourier transforms we  get the original wavefunction $s(k_0,...,k_{n-1})$.
 For example from ${\widetilde s}(j_0,...,j_{n-1})$ we go to $s_{n-1}(j_0,...,j_{n-2}|k_0)]$ as follows:
 \begin{eqnarray}
 s_{n-1}(j_0,...,j_{n-2}|k_0)]=[\omega_{d^2}(-j_{n-2}k_0).... \omega_{d^n}(-j_0k_0)]\frac{1}{\sqrt{d}}\sum _{j_0}\omega_{d}(-j_{n-1}k_{0})
 {\widetilde s}(j_0,...,j_{n-1})  \end{eqnarray}
 In a similar way we go backwards in all above steps.
 Therefore all the $s_r(j_0,...,j_{r-1}|k_0,..,k_{n-r-1})$ contain the same information as the original wavefunction $s(k_0,...,k_{n-1})$.

 \end{itemize}
 \begin{example}
 
 For $n=2$, Eq.(\ref{99A}) becomes
\begin{eqnarray}
 {F}(j_0,j_1|k_0,k_1)=\frac{1}{\sqrt{d^2}}\omega _{d^2}[j_0k_0+d(j_1k_0+k_1j_0)].
 \end{eqnarray}
Acting on a vector $s(K)=s(k_0,k_1)$ we get
\begin{eqnarray}\label{500}
 {\widetilde s}(J)= {\widetilde s}(j_0,j_1)=\frac{1}{\sqrt{d^2}}\sum_{k_0,k_1}\omega _{d^2}[j_0k_0+d(j_1k_0+k_1j_0)]s(k_0,k_1).
 \end{eqnarray}
In this case the fast Fourier transform given above becomes
\begin{eqnarray}\label{501}
 {\widetilde s}(J)= {\widetilde s}(j_0,j_1)&=\frac{1}{\sqrt{d}}&\sum _{k_0}\omega_{d}(j_1k_{0})[\omega_{d^2}(j_{0}k_0)s_{1}(j_0|k_0)]
  \end{eqnarray}
  with
 \begin{eqnarray}\label{502}
 s_1(j_0|k_0)&=&\frac{1}{\sqrt{d}}\sum _{k_{1}}\omega_d(j_0k_1)s(k_0,k_1)
 \end{eqnarray}

 \end{example}
 
 \subsection{Limitation of the method}\label{R}
 We have calculated the Fourier transform of the function $s(K)=s( k_0,...,k_{n-1})$.
 For other functions it is not easy to apply this method.
 For example in the Weyl function in Eq.(\ref{WW}) below, we want to calculate the Fourier transform of the function $s(K)s^*(B+K)$.
 Because the rings $[{\mathbb Z}(d)]^n$ and ${\mathbb Z}(D)$ (with $D=d^n$) are not isomorphic to each other, if 
 \begin{eqnarray}
&&(k_0,...,k_{d-1})\;\leftrightarrow\;K=k_0+k_1d+...+k_{n-1}d^{n-1}\nonumber\\
&&(b_0,...,b_{d-1})\;\leftrightarrow\;B=b_0+b_1d+...+b_{n-1}d^{n-1}.
\end{eqnarray}
the $(k_0+b_0,...,k_{n-1}+b_{n-1})$ does not correspond to $K+B$.
It is then difficult to apply directly the above formalism to Eq.(\ref{WW}) for the fast calculation of the Weyl  and Wigner functions.

In general, this fast Fourier transform is not directly applicable to functions which involve various sums and products of the variables.
The fact that the rings $[{\mathbb Z}(d)]^n$ and ${\mathbb Z}(D)$ are not isomorphic to each other, limits the practical use of the method.

 \subsection{Parallelism in the special case of factorisable states}\label{GF} 
 
 We consider the factorisable state
 \begin{eqnarray}\label{FFF}
 s(K)=s(k_0,...,k_{n-1})=g_0(k_0)g_1(k_1)...g_{n-1}(k_{n-1});\;\;\;\sum_{k_\nu}|g_\nu(k_\nu)|^2=1.
 \end{eqnarray}  
 In this case
  \begin{eqnarray}
 &&s_1(j_0|k_0,..,k_{n-2})=g_0(k_0)...g_{n-2}(k_{n-2}){\widetilde g}_{n-1}(j_0)\nonumber\\
 &&{\widetilde g}_{n-1}(j_0)=\frac{1}{\sqrt{d}}\sum _{k_{n-1}}\omega_d(j_0k_{n-1})g_{n-1}(k_{n-1})
    \end{eqnarray}
 Also
 \begin{eqnarray}
 &&s_2(j_0,j_1|k_0,..,k_{n-3})=g_0(k_0)...g_{n-3}(k_{n-3}){\widetilde G}_{n-2}(j_0,j_1){\widetilde g}_{n-1}(j_0)\nonumber\\
 &&{\widetilde G}_{n-2}(j_0,j_1)= \frac{1}{\sqrt{d}}\sum _{k_{n-2}}\omega_{d}(j_1k_{n-2}) [\omega_{d^2}(j_0k_{n-2})g_{n-2}(k_{n-2})]
 \end{eqnarray}  
Also
 \begin{eqnarray}
 &&s_3(j_0,j_1,j_2|k_0,..,k_{n-4})=g_0(k_0)...g_{n-4}(k_{n-4}){\widetilde G}_{n-3}(j_0,j_1,j_2){\widetilde G}_{n-2}(j_0,j_1){\widetilde g}_{n-1}(j_0)\nonumber\\
 &&{\widetilde G}_{n-3}(j_0,j_1,j_2)= \frac{1}{\sqrt{d}}\sum _{k_{n-3}}\omega_{d}(j_2k_{n-3}) [\omega_{d^2}(j_1k_{n-3})\omega_{d^3}(j_0k_{n-3})g_{n-3}(k_{n-3})]
 \end{eqnarray}  
  etc. The last one is
   \begin{eqnarray}\label{991}
&& {\widetilde s}(j_0,...,j_{n-1})={\widetilde G}_0(j_0,...,j_{n-1}){\widetilde G}_1(j_0,...,j_{n-2})...{\widetilde G}_{n-2}(j_0,j_1){\widetilde g}_{n-1}(j_0)\nonumber\\
 &&{\widetilde G}_0(j_0,...,j_{n-1})=\frac{1}{\sqrt{d}}\sum _{k_0}\omega_{d}(j_{n-1}k_{0})[\omega_{d^2}(j_{n-2}k_0).... \omega_{d^n}(j_0k_0)g_0(k_0)]
  \end{eqnarray}   
 We note that for factorisable functions, we can calculate independently each of the $n$ factors ${\widetilde G}_0(j_0,...,j_{n-1}), {\widetilde G}_1(j_0,...,j_{n-2}),...,{\widetilde g}_{n-1}(j_0)$
 and multiply them at the end. Therefore this scheme is suitable for parallel computation. 
 The calculation in the previous subsection for general functions, needs to be done sequentially.

 \begin{example}
 
For $n=2$ we consider the factorisable state
 \begin{eqnarray}
s(K)= s(k_0,k_1)=g_0(k_0)g_1(k_1)
 \end{eqnarray}  
 In this case
   \begin{eqnarray}\label{991}
&&{\widetilde s}(J)={\widetilde s}(j_0,j_1)={\widetilde G}_0(j_0,j_1){\widetilde g}_{1}(j_0)\nonumber\\
 &&{\widetilde g}_{1}(j_0)=\frac{1}{\sqrt{d}}\sum _{k_1}\omega_d(j_0k_1)g_1(k_1)\nonumber\\
  &&{\widetilde G}_0(j_0,j_1)=\frac{1}{\sqrt{d}}\sum _{k_0}\omega_{d}(j_{1}k_{0})[\omega_{d^2}(j_{0}k_0)g_0(k_0)]
  \end{eqnarray}   
  The two factors ${\widetilde G}_0(j_0,j_1)$ and ${\widetilde g}_{1}(j_0)$ can be calculated in parallel.

\end{example}
 \begin{remark}
 The `parallel formalism' of this section is limited to special cases where we know that the factorisation in Eq.(\ref{FFF}) holds.
 Given $s(K)=s(k_0,...k_{n-1})$, we give a necessary (but not sufficient) condition for the factorisation to hold.
 
 We define the  
  \begin{eqnarray}
|g(k_\nu)|^2= \sum _{\ne k_\nu}|s(k_0,...k_{n-1})|^2.
 \end{eqnarray}  
 Here we have a summation over all indices, except one.
 A necessary (but not sufficient) condition for Eq.(\ref{FFF}) to hold, is that   
 \begin{eqnarray}
|s(k_0,...k_{n-1})|=|g(k_0)|...|g(k_{n-1})|
 \end{eqnarray}  
 
 \end{remark}

 \subsection{Time complexity of the Fourier transform: counting the number of multiplications}\label{CC}

The estimate of the computational time is usually based on the number of multiplications, because they require more computational time than additions.
It is easily seen that the number of multiplications for `normal' Fourier transform is ${\mathcal O}(D^2)$ (it is a multiplication of a $D\times D$ matrix with a $D$-dimensional vector).
For the fast Fourier transform it is known that a lower bound for the computational time is ${\mathcal O}(D\log D)$, and we now give an approximate estimate for this.

In the fast transform in section \ref{sec10}, the first step in Eq.(\ref{20}) is a Fourier transform in a $d$-dimensional space and it requires $d^2$ multiplications.
This needs to be repeated for all values of the $n-1$ variables $k_0,..,k_{n-2}$ which take $d$ values each, therefore the number of multiplications is $d^2d^{n-1}=Dd$. 
The second step in Eq.(\ref{20A}) involves another $Dd$ multiplications (plus some extra multiplications which we ignore because we are interested in a lower limit).
In this way we find that a lower bound for the number of multiplications is 
 \begin{eqnarray}
Dnd\ge Dn\log d=D\log D.
 \end{eqnarray}
Many authors pointed out that this is a lower bound and that `real' numerical fast Fourier transforms take a bit more time than that.

\begin{figure}
\centering
\includegraphics{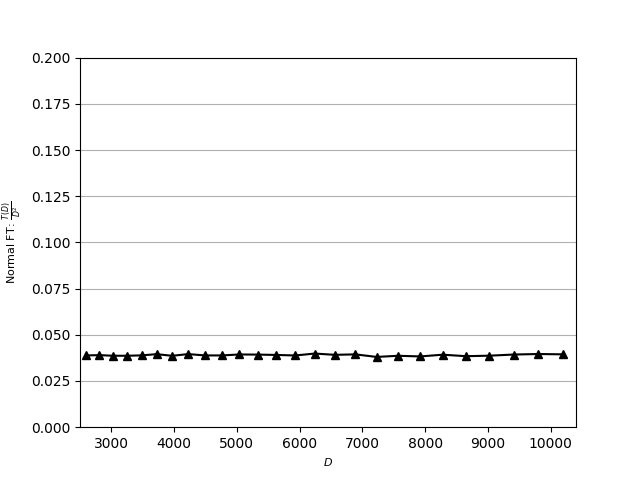}
\caption{$\frac{T(D)}{D^2}$ for the `normal' Fourier transform of a random vector (Eq(\ref{500})), as a function of $D=d^2$ where $d=51,53,...,101$ is an odd integer.
The result is a horizontal line, and this confirms that  
the computational time for the normal Fourier transform is ${\mathcal O}(D^2)$.
The Fourier transform of different random vectors give similar results.
}
\label{Fig2}
\end{figure}

\begin{figure}
\centering
\includegraphics{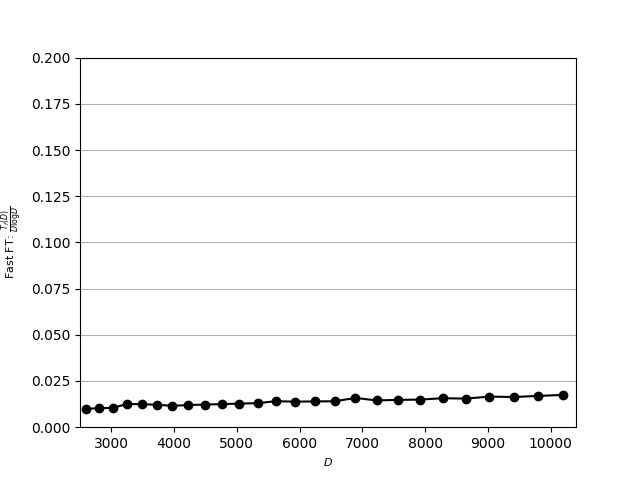}
\caption{
$\frac{T_f(D)}{D\log D}$ for the fast Fourier transform of a random vector(Eqs(\ref{501}), (\ref{502})), as a function of $D=d^2$ where $d=51,53,...,101$ is an odd integer. The result is a slightly ascending line, and this confirms that  
a lower bound for the computational time for the fast Fourier transform is approximately ${\mathcal O}(D\log D)$.
The Fourier transform of different random vectors give similar results.}
\label{Fig3}
\end{figure}

We consider a Hilbert space $H(D)$ with $D=d^2$,  where $d$ that takes all the odd values $51,....,101$. 
Using a random vector $s(K)=s(k_0,k_1)$ (produced by qiskit \cite{qiskit}) , we calculated ${\widetilde s}(J)={\widetilde s}(j_0,j_1)$ using both Eq(\ref{500}) 
(that involves the multiplication of a $D\times D$ matrix times a $D$-dimensional vector) and 
also the fast Fourier transform in Eqs(\ref{501}), (\ref{502}). 
We call $T(D)$ the computational time for the calculation of all components ${\widetilde s}(j_0,j_1)$ with the `normal' Fourier transform in Eq.(\ref{500}), and $T_f(D)$ the computational time for the calculation with the fast Fourier transform
in Eqs(\ref{501}), (\ref{502}).
In Figs.\ref{Fig2},\ref{Fig3} we plot 
 \begin{eqnarray}
\frac{T(D)}{D^2};\;\; \frac{T_f(D)}{D\log D};\;\;\;D=d^2.
 \end{eqnarray}
 The result for $\frac{T(D)}{D^2}$ in Fig.\ref{Fig2} is a horizontal line 
  and this confirms that  
the computational time for the normal Fourier transform is ${\mathcal O}(D^2)$. 

The result for $\frac{T_f(D)}{D\log D}$ in Fig.\ref{Fig3} is a slightly ascending  line and this confirms that a good lower bound for
the computational time of the fast Fourier transform is approximately ${\mathcal O}(D\log D)$.

\begin{figure}
\centering
\includegraphics{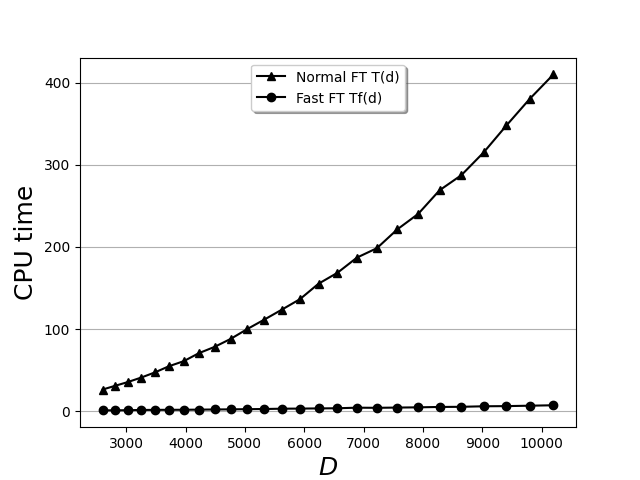}
\caption{The CPU times $T(D)$ and $T_f(D)$ for the normal Fourier transform (Eq(\ref{500})) and the fast Fourier transform (Eqs(\ref{501}), (\ref{502})) correspondingly, of a random vector, as a function of $D=d^2$ 
where $d=51,53,...,101$ is an odd integer.
It is seen that $T_f(D)$ is much smaller than $T(D)$. Different random vectors give similar results.}

\label{Fig4}
\end{figure}

\begin{figure}
\centering
\includegraphics[scale =0.4]{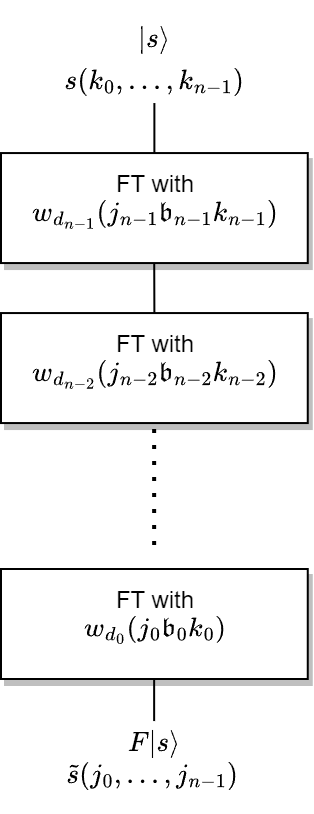}
 \caption{Circuit for the fast implementation of the Fourier transform$f$ using Eqs(\ref{A1})-(\ref{A2}). Here $D=d_0...d_{n-1}$ with coprime $d_0,...,d_{n-1}$. The constants ${\mathfrak b}_\nu$ are defined in Eq.(\ref{7}).}

\label{Fig5}
\end{figure}

In Fig.\ref{Fig4} we compare $T(D)$ with $T_f(D)$. It is seen that $T_f(D)$ is much smaller than $T(D)$.
We checked that the Fourier transform of different random vectors give similar results.

\section{The case $D=d_0...d_{n-1}$ with coprime $d_0,...,d_{n-1}$}

\subsection{A multipartite system $\Sigma(d_0,...,d_{n-1})$ with variables in ${\mathbb Z}(d_0)\times...\times{\mathbb Z}(d_{n-1})$. }

In this section $D=d_0...d_{n-1}$ with $d_0,...,d_{n-1}$ odd integers coprime to each other.
We consider a multipartite system  $\Sigma(d_0,...,d_{n-1})$ comprised of $n$ components, which are described with variables in ${\mathbb Z}(d_0),...,{\mathbb Z}(d_{n-1})$.
Positions and momenta in the multipartite system take values in ${\mathbb Z}(d_0)\times...\times {\mathbb Z}(d_{n-1})$ and the corresponding  Hilbert space is ${\mathfrak H}_B=H(d_0)\otimes ...\otimes H(d_{n-1})$.
The Hilbert spaces $H(D)$ and ${\mathfrak H}_B$ are isomorphic (they have the same dimension), and therefore $\Sigma (D)$ and  $\Sigma(d_0,...,d_{n-1})$  are two different descriptions of the same system.

We consider the basis
\begin{eqnarray}
\ket{X;j_0,...,j_{n-1}}=\ket{X;j_0}\otimes...\otimes\ket{X;j_{n-1}};\;\;\;j_\nu\in{\mathbb Z}(d_\nu).
\end{eqnarray}
An arbitrary state is written as
\begin{eqnarray}\label{5}
\ket{s}=\sum s(j_0,...,j_{n-1})\ket{X;j_0,...,j_{n-1}};\;\;\;\sum |s(j_0,...,j_{n-1})|^2=1.
\end{eqnarray}

Fourier transforms in $\Sigma(d_0,...,d_{n-1})$ are defined as:
\begin{eqnarray}
&&{\mathfrak F}_B={\cal F}_0\otimes ...\otimes {\cal F}_{n-1};\;\;\;{\cal F}_\nu=\frac{1}{\sqrt{d}}\sum _{j_\nu ,k_\nu}\omega_{d_{\nu}}(j_{\nu} k_{\nu}) \ket{X;j_{\nu}}\bra{X;k_{\nu}}\nonumber\\
&&{\mathfrak F}_B^4={\bf 1};\;\;\;{\mathfrak F}_B{\mathfrak F}_B^{\dagger}={\bf 1};\;\;\;j_\nu,k_\nu\in{\mathbb Z}(d_\nu).
\end{eqnarray}
Clearly $F$ is very different from ${\mathfrak F}_B$.

\subsection{Fast Fourier transform $F$ in $\Sigma(D)$ as a sequence of transformations in $\Sigma(d_0,...,d_{n-1})$ with $D=d_0...d_{n-1}$}\label{sec11}

We use the following  dual notation for all functions and states in $\Sigma(D)$, based on the bijective map in Eq.(\ref{9}):
\begin{eqnarray}\label{234}
s(K)=s(k_0,...,k_{n-1});\;\;\;k_\nu\in {\mathbb Z}(d_\nu).
\end{eqnarray}

Using Eq.(\ref{13A}) we express the matrix elements of the Fourier transform $F$ in $\Sigma(D)$ (Eq.(\ref{FF})) as:
\begin{eqnarray}\label{99}
 &&F(j_0,...,j_{n-1}|k_0,...,k_{n-1})=\bra{j_0,...,j_{n-1}}F\ket{k_0,...,k_{n-1}}\nonumber\\
  &&=\left [\frac{1}{\sqrt{d_0}}\omega_{d_0}( j_0{\mathfrak b}_0 k_0)\right ]...\left [\frac{1}{\sqrt{d_{n-1}}}\omega_{d_{n-1}}( j _{n-1}{\mathfrak b}_{n-1} k_{n-1})\right ]
   \end{eqnarray}
   The constants ${\mathfrak b}_\nu$ have been defined in Eq.(\ref{7}).
 Using this we implement the Fourier transform in Eq.(\ref{TR}), as a sequence of transforms in the system $\Sigma(d_0,...,d_{n-1})$.
 It involves the following steps (shown also in the quantum circuit in Fig.\ref{Fig5}):
  \begin{itemize}
 \item
 A Fourier transform of $s(K)=s(k_0,...,k_{n-1})$ with $\omega_{d_{n-1}}( j _{n-1}{\mathfrak b}_{n-1} k_{n-1})$ (we note here the constant ${\mathfrak b}_{n-1}$) and summation over $k_{n-1}$:
 \begin{eqnarray}\label{A1}
 s_1(j_{n-1}|k_0,..,k_{n-2})=\frac{1}{\sqrt{d_{n-1}}}\sum _{k_{n-1}}\omega_{d_{n-1}}( j _{n-1}{\mathfrak b}_{n-1} k_{n-1})s(k_0,...,k_{n-1})
   \end{eqnarray}
\item
A Fourier transform of $s_1(j_{n-1}|k_0,...,k_{n-2})$ with $\omega_{d_{n-2}}( j _{n-2}{\mathfrak b}_{n-2} k_{n-2})$ (we note here the constant ${\mathfrak b}_{n-2}$) and summation over $k_{n-2}$:
 \begin{eqnarray}\label{AA1}
 s_2(j_{n-2}, j_{n-1}|k_0,..,k_{n-3})=\frac{1}{\sqrt{d_{n-2}}}\sum _{k_{n-2}}\omega_{d_{n-2}}( j _{n-2}{\mathfrak b}_{n-2} k_{n-2}) s_1(j_{n-1}|k_0,..,k_{n-2}),
   \end{eqnarray}
etc. The last step is
\item
  \begin{eqnarray}\label{A2}
{\widetilde s}(J)={\widetilde s}(j_0,...,j_{n-1})=\frac{1}{\sqrt{d_0}}\sum _{k_0}\omega_{d_0}( j _0{\mathfrak b}_0 k_0)s_{n-1}(j_1,...,j_{n-1}|k_0)
  \end{eqnarray}   
Similarly to the previous method, for factorisable functions these $n$ steps can be done in parallel.
But for general functions, they need to be done sequentially.
\end{itemize}

 \subsection{Time complexity of the Fourier transform: counting the number of multiplications}

We first give an approximate estimate that a lower bound for the computational time in the present scheme, is ${\mathcal O}(D\log D)$.

In the fast transform in section \ref{sec11}, the first step in Eq.(\ref{A1}) is a Fourier transform in a $d_{n-1}$-dimensional space and it requires $d_{n-1}^2$ multiplications.
This needs to be repeated for all values of the $n-1$ variables $k_0,..,k_{n-2}$, therefore the number of multiplications is $d_1...d_{n-2}d_{n-1}^2=Dd_{n-1}$. 
The second step in Eq.(\ref{AA1}) involves another $Dd_{n-2}$ multiplications.
In this way we find that a lower bound for the number of multiplications is 
 \begin{eqnarray}
D(d_0+...+d_{n-1})\ge D(\log d_0+...+\log d_{n-1})=D\log D.
 \end{eqnarray}

\begin{figure}
\centering
\includegraphics{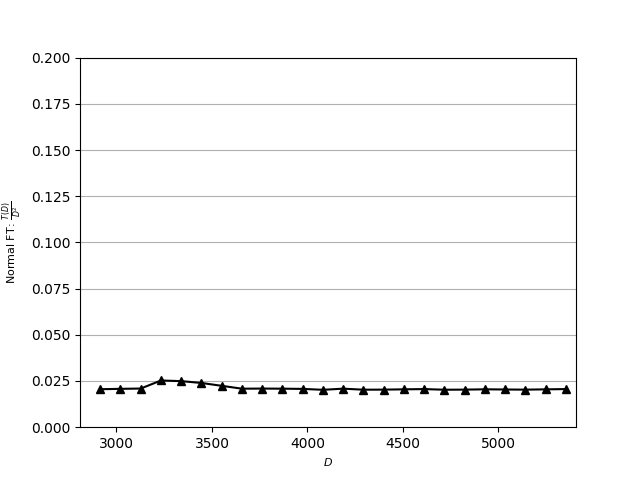}
\caption{$\frac{T(D)}{D^2}$ for the `normal' Fourier transform of a random vector (Eq(\ref{500})), as a function of $D=d_1d_2$ where $d_1=53$ and $d_2=55,57,...,101$ (the $d_1,d_2$ are coprime).
The result is a horizontal line, and this confirms that  
the computational time for the normal Fourier transform is ${\mathcal O}(D^2)$.
The Fourier transform of different random vectors give similar results.
}
\label{Fig6}
\end{figure}

\begin{figure}
\centering
\includegraphics{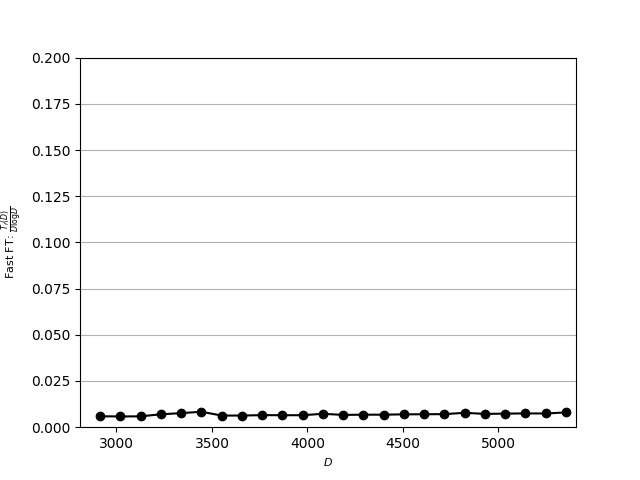}
\caption{
$\frac{T_f(D)}{D\log D}$ for the fast Fourier transform of a random vector(Eqs(\ref{501}), (\ref{502})), as a function of $D=d_1d_2$ where $d_1=53$ and and $d_2=55,57,...,101$ (the $d_1,d_2$ are coprime).
The result is approximately a horizontal line, and this confirms that  
the computational time for the fast Fourier transform is approximately ${\mathcal O}(D\log D)$.
The Fourier transform of different random vectors give similar results.}
\label{Fig7}
\end{figure}

We consider Hilbert spaces $H(d_1d_2)$ where $d_1=53$ and $d_2$ takes the odd values $55, 57,....,101$. Since $53$ is a prime number the $d_1, d_2$ are coprime.
As in section \ref{CC} we used a random vector $s(K)=s(k_0,k_1)$ (produced by qiskit \cite{qiskit}) , we calculated ${\widetilde s}(J)={\widetilde s}(j_0,j_1)$.
In Figs.\ref{Fig6},\ref{Fig7} we plot 
 \begin{eqnarray}
\frac{T(D)}{D^2};\;\; \frac{T_f(D)}{D\log D};\;\;\;D=d_1d_2.
 \end{eqnarray}
 The result for $\frac{T(D)}{D^2}$ in Fig.\ref{Fig6} is a horizontal line and this confirms that  
the computational time for the normal Fourier transform is ${\mathcal O}(D^2)$.
  The result for $\frac{T_f(D)}{D\log D}$ in Fig.\ref{Fig7} is also a horizontal  line and this confirms that a good lower bound for
the computational time of the fast Fourier transform is approximately ${\mathcal O}(D\log D)$.

\begin{figure}
\centering
\includegraphics{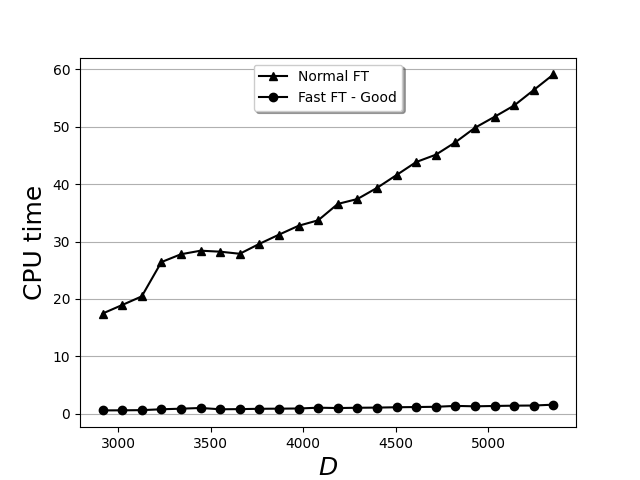}
\caption{The CPU times $T(D)$ and $T_f(D)$ for the normal Fourier transform (Eq(\ref{500})) and the fast Fourier transform (Eqs(\ref{501}), (\ref{502})) correspondingly, of a random vector, 
as a function of $D=d_1d_2$ where $d_1=53$ and and $d_2=55,57,...,101$ (the $d_1,d_2$ are coprime).
It is seen that $T_f(D)$ is much smaller than $T(D)$. Different random vectors give similar results.}

\label{Fig8}
\end{figure}

In Fig.\ref{Fig8} we compare $T(D)$ with $T_f(D)$. It is seen that $T_f(D)$ is much smaller than $T(D)$.
We checked that the Fourier transform of different random vectors give similar results.

\section{Fast Wigner and Weyl functions using the second method}

Phase space methods for the system $\Sigma (D)$ (Wigner and Weyl functions, etc) rely heavily on Fourier transforms.
Therefore fast Fourier transforms can be used for the fast calculation of various quantities within the phase space formalism.

As an example, we consider  the Weyl function $\widetilde W(A,B)$ and the Wigner function $W(A,B)$ for the state $\ket{s}=\sum _Ks(K)\ket{X;  K}$ of the system $\Sigma(D)$.
They are given by the following Fourier transforms(e.g.,\cite{VV}):
\begin{eqnarray}\label{WW}
&&\widetilde W(A,B)=\omega_D(2^{-1}AB)\sum _K \omega_D(AK)s(K)s^*(B+K);\;\;\;A,B\in{\mathbb Z}(D)\nonumber\\
&&W(A,B)=\omega_D(2AB)\sum _K \omega_D(-2AK)s(K)s^*(2B-K)
\end{eqnarray}
The $2^{-1}=\frac{D+1}{2}({\rm mod}\;D)$ for odd $D$.

We explained in subsection \ref{R} that the first method for fast Fourier transforms (in the case $D=d^n$) is not directly applicable to Eqs(\ref{WW}), for the fast calculation of these functions.
This is related to the fact that the rings $[{\mathbb Z}(d)]^n$ and ${\mathbb Z}(D)$ (with $D=d^n$) are not isomorphic to each other.

\begin{figure}
\centering
\includegraphics[scale =0.4]{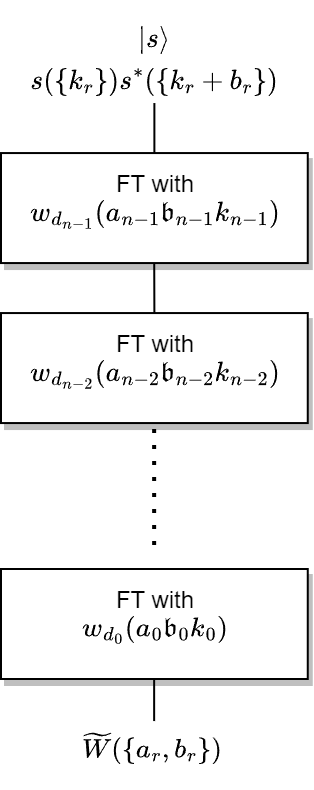}
 \caption{Circuit for the fast calculation of the Weyl function, using Eqs(\ref{S1})-(\ref{S3}). Here $D=d_0...d_{n-1}$ with coprime $d_0,...,d_{n-1}$. The constants ${\mathfrak b}_\nu$ are defined in Eq.(\ref{7}).}

\label{Fig9}
\end{figure}

The second method for fast Fourier transforms (in the case $D=d_0...d_{n-1}$ with coprime $d_0,...,d_{n-1}$) is directly applicable in the fast calculation of the Weyl and Wigner functions.
We present in detail the fast Weyl function.
We use the bijective map in Eq.(\ref{9}), and express $K,B$ as
\begin{eqnarray}\label{234}
&&K\leftrightarrow(k_0,...,k_{n-1});\;\;\;k_\nu\in {\mathbb Z}(d_\nu)\nonumber\\
&&B\leftrightarrow(b_0,...,b_{n-1});\;\;\;b_\nu\in {\mathbb Z}(d_\nu)\nonumber\\
&&A\leftrightarrow(a_0,...,a_{n-1});\;\;\;a_\nu\in {\mathbb Z}(d_\nu)
\end{eqnarray}
The rings ${\mathbb Z}(D)$ and ${\mathbb Z}(d_0)\times ...\times {\mathbb Z}(d_{n-1})$ are isomorphic and therefore
\begin{eqnarray}\label{234}
K+B\leftrightarrow(k_0+b_0,...,k_{n-1}+b_{n-1}).
\end{eqnarray}
Consequently
\begin{eqnarray}
s(K)s^*(B+K)=s(k_0,...,k_{n-1})s^*(k_0+b_0,...,k_{n-1}+b_{n-1}).
\end{eqnarray}
We now give briefly the basic steps for the fast Weyl function (shown also in the quantum circuit in Fig\ref{Fig9}). 
\begin{itemize}
 \item
 A Fourier transform of $s(\{k_{r}\})s^*(\{k_r+b_r\})$ with $\omega_{d_{n-1}}( a_{n-1}{\mathfrak b}_{n-1} k_{n-1})$  (we note here the constant ${\mathfrak b}_{n-1}$) and summation over $k_{n-1}$:
 \begin{eqnarray}\label{S1}
 {\widetilde W}_1(a_{n-1}|k_0,..,k_{n-2}|\{b_r\})=\sum _{k_{n-1}}\omega_{d_{n-1}}( a_{n-1}{\mathfrak b}_{n-1} k_{n-1})s(\{k_r\})s^*(\{k_r+b_r\}).
    \end{eqnarray}
\item
A Fourier transform of $ {\widetilde W}_1(a_{n-1}|k_0,..,k_{n-2}|\{b_r\})$ with $\omega_{d_{n-2}}( a _{n-2}{\mathfrak b}_{n-2} k_{n-2})$ (we note here the constant ${\mathfrak b}_{n-2}$) and summation over $k_{n-2}$:
 \begin{eqnarray}\label{S2}
 {\widetilde W}_2(a_{n-2}, a_{n-1}|k_0,..,k_{n-3}|\{b_r\})=\sum _{k_{n-2}}\omega_{d_{n-2}}( a_{n-2}{\mathfrak b}_{n-2} k_{n-2})  {\widetilde W}_1(a_{n-1}|k_0,..,k_{n-2}|\{b_r\}),
   \end{eqnarray}
etc. The last step is
\item
  \begin{eqnarray}\label{S3}
{\widetilde W}(A,B)={\widetilde W}(\{a_r,b_r\})=\omega_D(2^{-1}AB)\sum _{k_0}\omega_{d_0}( a_0{\mathfrak b}_0 k_0){\widetilde W}_{n-1}(a_1,..., a_{n-1}|k_0|\{b_r\}).
 \end{eqnarray}   
Similarly to the previous methods, for factorisable functions $s(k_0,...,k_{n-1})$ these $n$ steps can be done in parallel.
But for general functions, they need to be done sequentially.
\end{itemize}
Analogous algorithm can be given for the Wigner function.

We note that the Fourier transform requires ${\cal O}(D^2)$ multiplications, but it needs to be performed for all values of $A,B$.
Therefore the complexity of the calculation of the Wigner or Weyl function is ${\cal O}(D^4)$, and with the fast Fourier transforms discussed above it is reduced to ${\cal O}(D^3\log D)$.

As an example we consider the case $D=21\times 23=3\times 7\times 23$ and calculated the Weyl function of a random vector (produced by qiskit\cite{qiskit}) with the normal Fourier transform and with the fast method given above.
We found numerically that the ratio of the corresponding computational times is $T/T_f=14.7$ (with the $D=21\times 23$ factorisation), and $T/T_f=17.6$ (with the $D=3\times 7\times 23$ factorisation).

\section{Discussion}

We have presented a fast implementation of the Fourier transform $F$ in a large quantum system.
This replaces the large Fourier transform with many small Fourier transforms.
The small Fourier transforms can be performed classically or (if available) in a quantum computer in which case we have the well known additional advantages of quantum Fourier transforms.
We used two methods.

The first method is for the case $D=d^n$ with $d$ an odd integer. This is based on the bijective map between the sets ${\mathbb Z}(D)$ and $[{\mathbb Z}(d)]^n$ in Eq.(\ref{16}).
The algorithm is described in Eqs(\ref{20})-(\ref{30}) and the relevant quantum circuit is shown in Fig.\ref{Fig1}.

The complexity (based on the number of multiplications) of the normal Fourier transform  is  ${\cal O}(D^2)$ and of the fast Fourier transform ${\cal O}(D\log D)$.
This has been supported with numerical work shown in figs \ref{Fig2},\ref{Fig3}.
As expected the fast Fourier transform is much faster than the normal Fourier transform (Fig.\ref{Fig3}). 
A limitation of the method is the fact that the ring ${\mathbb Z}(D)$ (with $D=d^n$) is not isomorphic to the ring $[{\mathbb Z}(d)]^n$.
Consequently, this method cannot be used with Eqs(\ref{WW}) for the fast calculation of the Wigner and Weyl functions.

The second method is for the case $D=d_0...d_{n-1}$ with $d_0,...,d_{n-1}$ odd integers coprime to each other.
This is based on the bijective map between the rings ${\mathbb Z}(D)$ and  ${\mathbb Z}(d_0),...,{\mathbb Z}(d_{n-1})$ in Eq.(\ref{9}).
These two rings are isomorphic.
The algorithm is described in Eqs(\ref{A1})-(\ref{A2}) and the relevant quantum circuit is shown in Fig.\ref{Fig5}.
Numerical work shown in figs \ref{Fig6},\ref{Fig7} confirm that the complexity of the normal Fourier transform  is  ${\cal O}(D^2)$ and of the fast Fourier transform ${\cal O}(D\log D)$.
Fig\ref{Fig8} shows that the fast Fourier transform requires much less computational time than the Normal Fourier Transform.

This second method can be used with Eqs(\ref{WW}) for the fast calculation of the Wigner and Weyl functions.
The algorithm for the Weyl function is given in Eqs(\ref{S1})-(\ref{S3}) and the relevant quantum circuit is shown in Fig.\ref{Fig9}.

\section{Conflict of interest}
We have no conflict of interest to disclose.

\section{Data statement}
No data were used in this paper.

\end{document}